# Generation of GeV protons from 1 PW laser interaction with near critical density targets

Stepan S. Bulanov[1,2], Valery Yu. Bychenkov[3], Vladimir Chvykov[1], Galina Kalinchenko[1], Dale William Litzenberg[4], Takeshi Matsuoka[1], Alexander G. R. Thomas[1], Louise Willingale[1], Victor Yanovsky[1], Karl Krushelnick[1] and Anatoly Maksimchuk[1]

[1]FOCUS Center and Center for Ultrafast Optical Science, University of Michigan, Ann Arbor, Michigan 48109, USA

[2]Institute of Theoretical and Experimental Physics, Moscow 117218, Russia

[3]P. N. Lebedev Physics Institute, Russian Academy of Sciences, Moscow 119991, Russia

[4]Department of Radiation Oncology, University of Michigan, Ann Arbor, Michigan 48109, USA

Abstract

The propagation of ultra intense laser pulses through matter is connected with the generation of strong moving magnetic fields in the propagation channel as well as the formation of a thin ion filament along the axis of the channel. Upon exiting the plasma the magnetic field displaces the electrons at the back of the target, generating a quasistatic electric field that accelerates and collimates ions from the filament. Two-dimensional Particle-in-Cell simulations show that a 1 PW laser pulse tightly focused on a near-critical density target is able to accelerate protons up to an energy of 1.3 GeV. Scaling laws and optimal conditions for proton acceleration are established considering the energy depletion of the laser pulse.

PACS : 52.38.Kd, 29.25.Ni, 52.65.Rr,

## I. Introduction

The acceleration of charged particles from intense laser interactions with targets of different density and composition is considered to be one of the main applications of high power laser systems. In particular, the acceleration of ions has attracted a lot of interest over the last few years. The accelerated ions can potentially be used for fusion ignition [1], hadron therapy [2], radiography of dense targets [3], and injection into conventional accelerators [4]. Ion beams with a maximum energy of tens of MeV were observed in previous experiments from laser interactions with solid and gaseous targets [5,6]. The generation of energetic ions beams has also been thoroughly studied using both two dimensional (2D) and three dimensional (3D) particle-in-cell (PIC) computer simulations [7,8]. These simulations show that with laser systems capable of producing ultra-short pulses in the multiterawatt or even petawatt power range it is possible to generate ion beams with energies of several hundreds MeV or even several GeV by optimizing the parameters of the pulse and the target, and employing new regimes of acceleration.

There are several regimes of ion acceleration discussed in the literature: (i) Target Normal Sheath Acceleration (TNSA) [9], (ii) Coulomb Explosion (CE) [10], and (iii) Radiation Pressure or Laser Piston regime (LP) [11]. In these three regimes the laser pulse interacts with a thin foil of solid density. When the pulse is not intense enough to burn through the target, it launches hot electrons from the front surface through the target. Upon reaching the back of the target they establish a sheath electrostatic field which accelerates ions (TNSA). When the laser pulse is sufficiently intense, it effectively removes all the electrons from the irradiated volume and subsequently the bare ion core explodes due to the repulsion of noncompensated positive charges (CE). In the radiation pressure regime, the laser pulse is able to push the foil as whole due to the fact that as it is reflected it acts as a flying relativistic mirror (LP). Recently several new schemes were proposed, based on the enhancement and different combinations of these regimes: (i) Breakout afterburner – the effective combination of TNSA with the direct acceleration by the burning through laser pulse [12]; (ii) the enhanced Coulomb Explosion, where the ions are injected into a Coulomb field [13]; (iii) the Directed Coulomb explosion [14,15], which is achieved in laser double-layer foil interactions. In the Directed Coulomb Explosion regime the electrons are expelled from the focal spot, the first layer of heavy ions are accelerated by the radiation pressure, and then experience a Coulomb explosion, transforming into a positively charged cloud expanding in the direction of laser pulse propagation. The second layer ions are accelerated in the moving charge separation electric field of this cloud.

In this paper we report on the study of a different and potentially more efficient mechanism of ion acceleration. Whereas in all the above mentioned schemes thin and ultra-thin solid density targets are used, here we utilize a regime of laser pulse interaction with near critical density targets. These targets have the thickness which is larger than the laser pulse length. When laser propagates through such a targets, it forms a channel in both the electron and ion density. A portion of the electrons is accelerated in the direction of laser pulse propagation by the longitudinal electric field. The motion of these electrons generates a magnetic field, which is circulating in the channel around its axis. The region where the magnetic field is present moves behind the pulse. Upon exiting the channel the magnetic field expands into vacuum and the electron current is dissipated. This field has the form of a dipole in 2D and a toroidal vortex in 3D. The magnetic field displaces the electron component of plasma with regard to the ion component and a strong quasistatic electric field is generated that can accelerate and collimate ions. The accelerated ions originate from the thin ion filament that is formed along the axis of the propagation channel (see Fig. 1). The mechanism under consideration was proposed in Refs. [16] and studied in a number of papers [17,18,19]. It was shown recently that a short pulse can accelerate ions up to the maximum energy of 18 MeV per nucleon from cluster-gas targets [20]. In the case of long pulses the acceleration of helium ions up to 40 MeV from underdense plasmas was observed on the VULCAN laser [6]. However the scaling and optimal conditions were not established.

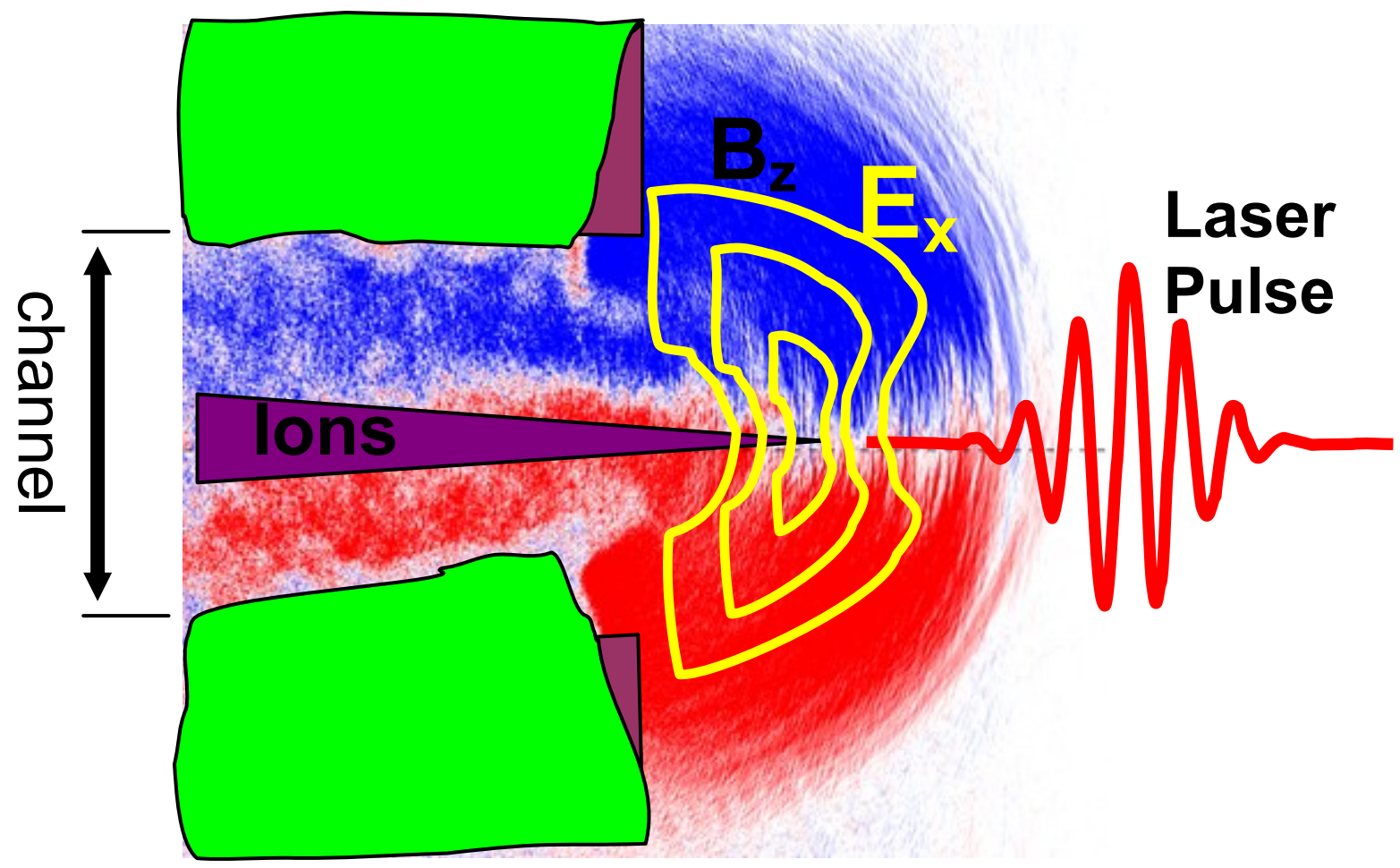

**Fig. 1. (color on-line)** The principal scheme of the acceleration mechanism.

We study the dependence of ion maximum energy on the target density, thickness as well as on the focusing and power of the laser pulse in order to optimize the acceleration process. We should note here the maximum proton energy is obtained for slightly overcritical targets and since the laser pulse should be able to establish a channel that goes through the target, the laser should be tightly focused on the front surface of the target to ensure the penetration through the target. We show that that it is not only important that the pulse penetrates the target, but also that it does not break into filaments. The latter will immediately reduce the effectiveness of acceleration. It is therefore necessary to establish matching between the dimensions of the focal spot, the position of the focus relative to the target boundary and the diameter of the self-focusing channel for each target density and thickness in order to avoid filamentation, as in was shown in Ref. [21]. The effectiveness of this mechanism depends on the efficient transfer of laser pulse energy into the energy of fast electrons that are accelerated along the propagation channel. Because of this, for each laser target configuration there exists an optimum target thickness that maximizes the ion energy. We show how the optimum target thickness scales with peak intensity and pulse duration. The energies of protons produced in such interaction by multi-Terawatt and Petawatt class lasers are of the order of several hundreds of MeV, which is interesting for many applications that require ion beams [1-4].

The paper is organized as follows. In section II we present the results of 2D PIC simulations of intense tightly focused laser pulse interaction with near critical density targets of different thickness and discuss the mechanism of proton acceleration. The scaling of the optimal target thickness with laser pulse parameters, based on the optimal laser pulse energy depletion, is presented in section III. We conclude in section IV.

## II. The results of 2D PIC simulations.

In this section we present the results of 2D PIC simulations of an intense laser pulse interaction with underdense and near-critical density targets. The simulations were performed using the REMP (Relativistic Electro Magnetic Particle) code [22]. Space and time are measured in units of laser pulse wavelength, $\lambda$, and wave period, $T=2\pi c/\omega$, correspondingly, $\omega$ is the laser pulse frequency. The grid mesh spacing is $\lambda/20$, and the time step is $T/40$. The total number of particles in the simulation box is about $5\text{x}10^6$. A laser pulse with Gaussian temporal and spatial profiles is introduced at the left boundary. The pulse duration is $\tau$=30 fs or $10\lambda$ ($1/e^2$) and it is focused to a $1.5\lambda$ (FWHM) spot size at a distance of $6\lambda$ from the left boundary (f/D=1.5) and the power is 1 PW ($I \sim 10^{23}$ W/cm$^2$). It is worth mentioning here that the simulations we performed for different f/D show that the optimal acceleration conditions are reached for f/D=1.5 ($E_p(400\ TW,\ 2.7n_{cr})=480\ MeV$), which we use throughout the paper, unless stated otherwise. For f/D=1 the pulse immediately goes into filamentation reducing the efficiency of acceleration, $E_p(400\ TW,\ 2.7n_{cr})=300\ MeV$. For f/D=3 the peak intensity drops leading to a reduction of maximum proton energy, $E_p(400\ TW,\ 2.7n_{cr})=370\ MeV$, meaning that thinner targets should be used.

At intensities of about $10^{23}$ W/cm$^2$ the effects of radiation backreaction can become important and lead to the modification of the laser-plasma interaction, as it was shown in Ref. [23]. Though we use peak intensities of about $10^{23}$ W/cm$^2$ in our 2D PIC simulations the effects of radiation backreaction are negligible in our case and are not taken into account. It is due to the fact that as it will be shown below the channel in the electron density is formed by the front of the pulse which expels the electron sideways. Thus when the maximum of the pulse arrives there is only a negligible amount of electrons or no electrons at all to interact with.

The target is $40\lambda$ wide, its left boundary is placed at $x=6\lambda$. Thus the laser is focused at the front of the target. Focusing the pulse before the target or inside the target leads either to reduced propagation length in plasma or to filamentation, which are both not beneficial to efficient proton acceleration. The target is composed of fully ionized hydrogen. The density is measured in units of the critical density, $n_{cr} = m_e\omega^2 / 4\pi e^2$, $m_e$ is the electron mass and $e$ is the unit charge. We vary the thickness and density of the target since these are two parameters the dependence on which we study. The highest density used in simulations is $16n_{cr}$.In view of grid mesh spacing $\lambda/20$, it gives 5 points per plasma wavelength. Most of the simulations were performed at $n_e=3n_{cr}$, which gives 12 points per plasma wavelength. It is enough to resolve the minimum characteristic scale of the problem.

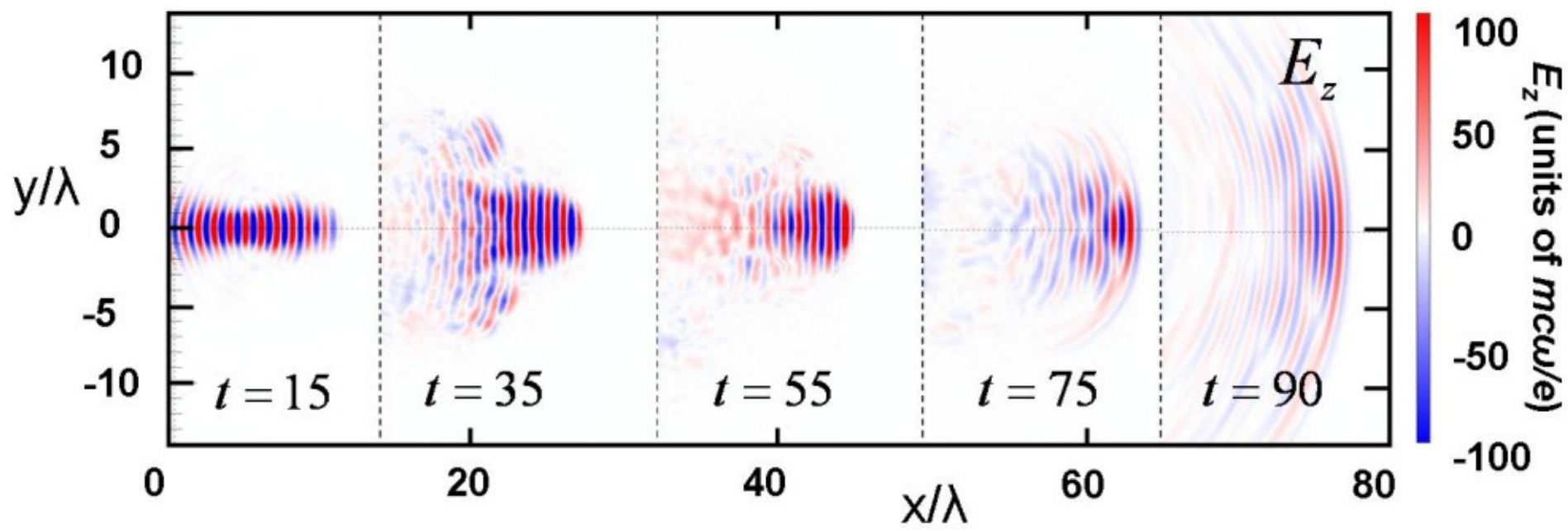

**Fig. 2. (color on-line)** The evolution of the laser pulse electric field for a 1 PW laser pulse interacting with a 50 λ thick target with a density of 3.0 $n_{cr}$.

When a tightly focused high-intensity laser pulse interacts with a target of near-critical density it forms a channel in both electron and ion density. The evolution of the laser field inside the channel is shown in Fig. 2. The pulse is tightly focused at the front surface of the target (t=15), after which it begins to diverge in the plasma and to expel the electrons creating a channel in electron density. The results of these simulations indicate that this process is connected with a swift rise of electron density on the walls of this channel. Soon the laser pulse divergence stops and the pulse propagates inside this channel (t=35, t=55), losing energy, which is transformed into the energy of fast electrons, which are mainly accelerated in the transverse direction. At t=75 the laser pulse exits the plasma and diverges (t=90). Comparing the pulse at t=15 and t=90 we see that a significant part of laser energy has been transferred to plasma electrons. The formation of the channel in electron density and a stream of accelerated electrons, which exit the plasma behind the pulse, along the laser propagation axis are shown in Fig. 3 a and b correspondingly. We notice here that the density in the stream of accelerated electrons is substantially higher than the electron density in the ambient plasma. In the equilibrium the electron density inside the bunch can be $\gamma_e^2$ times greater than the ion density in the plasma (e.g. see Ref. [23]). Here $\gamma_e$ is the electron bunch relativistic gamma-factor.

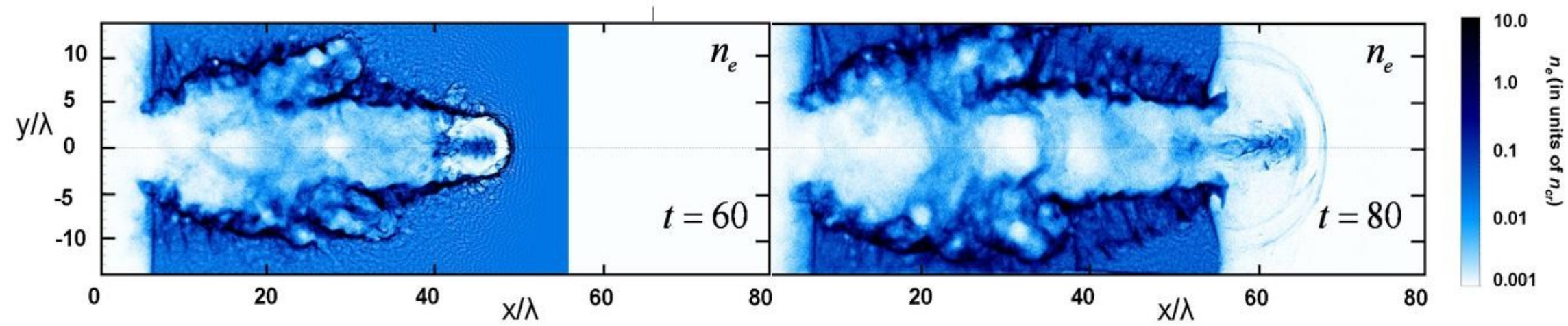

**Fig. 3. (color on-line) a)** The formation of a channel in electron density; **b)** The stream of laser accelerated electrons exiting plasma behind the pulse for a 1 PW laser pulse interacting with a 50 λ thick target with density of 3.0 $n_{cr}$.

These electrons generate a magnetic field, circulating inside the channel around its axis. The region, where the magnetic field is generated, moves behind the pulse. Upon exiting the plasma the magnetic field expands and the electron current is dissolved. Some of the electrons leave the target, other return back, helping to sustain the magnetic field on the back of the target (see Fig. 3b). This field displaces the electron component of plasma with regard to the ion component thus generating a strong quasistatic electric field that can accelerate and collimate ions. In order to illustrate the process of this electric field generation Fig. 4 shows the longitudinal electric and z-component of magnetic field for different moments of time. It can be clearly seen that the growth of the electric field is connected with the expansion of the magnetic field, which has a form of a dipole in 2D (in 3D it will be a toroidal vortex).

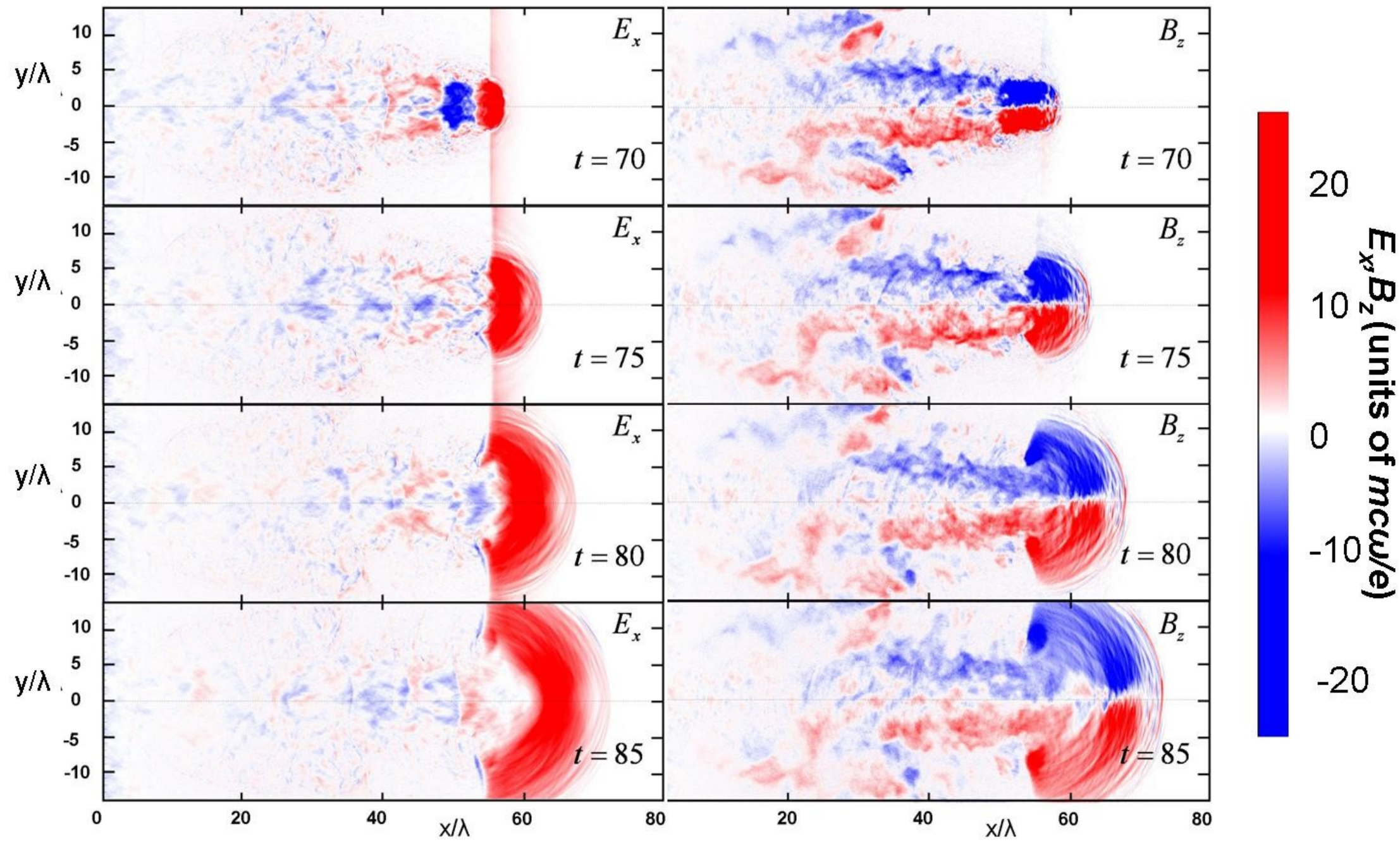

**Fig. 4. (Color on-line)** The evolution of the longitudinal electric and the z-component of the magnetic fields for a 1 PW laser pulse interacting with a 50 λ thick target with density of 3.0 $n_{cr}$.

This electric field will accelerate and collimate ions from the thin filament, which is formed along the laser propagation axis inside the channel in ion density. In Fig. 5 we present a series of proton density profiles at different time intervals to illustrate the creation of a channel, the formation of a thin proton filament and the proton acceleration from this filament by the longitudinal quasistatic electric field.

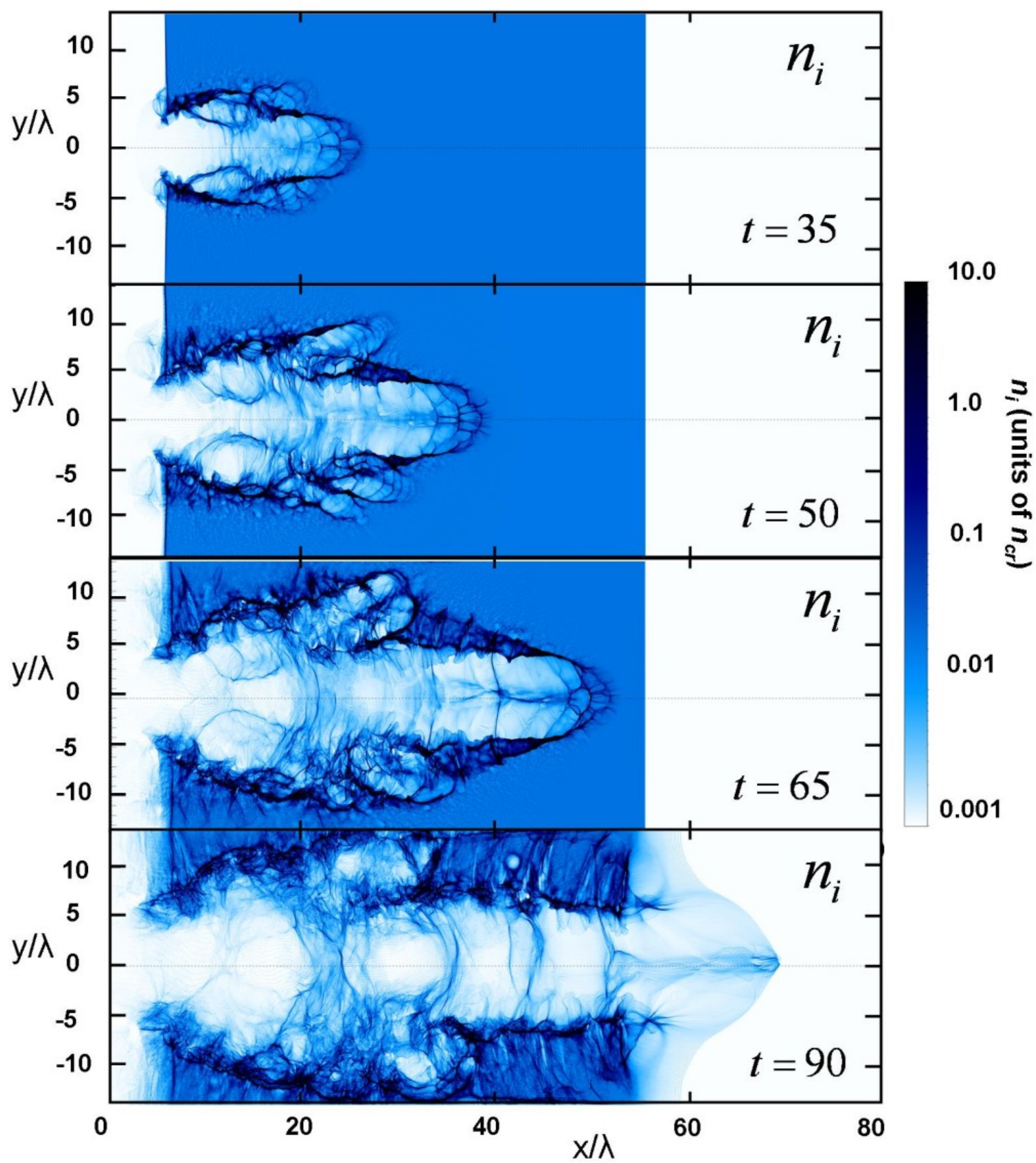

**Fig. 5. (Color on-line)** The evolution of ion density for a 1 PW laser pulse interacting with a 50 λ thick target with density of 3.0 $n_{cr}$.

In Fig. 6a we show a typical spectrum of ions for a 1 PW laser pulse interacting with a 50 λ thick target having a density of 3.0 $n_{cr}$ and producing protons with a maximum energy of 1.3 GeV. The number of protons with energy above 1 GeV is about $4 \times 10^8$. The dependence of ion maximum energy on time, shown in Fig. 6b, illustrates the acceleration mechanism. The steep rise in maximum energy begins at *t=70*, which corresponds to the formation of a quasistatic field at the back of the target (see Fig. 4). By *t=120* this field is almost dissolved and the acceleration stops. In Fig. 6c we present the angular distribution of protons. Two large side peaks at $\theta=\pi/2$ correspond to the protons pushed out from the channel by the laser pulse in the transverse direction. A narrow peak ($\Delta\theta=6^o$ at FWHM) at $\theta=0$ represents the protons accelerated along the laser propagation axis.

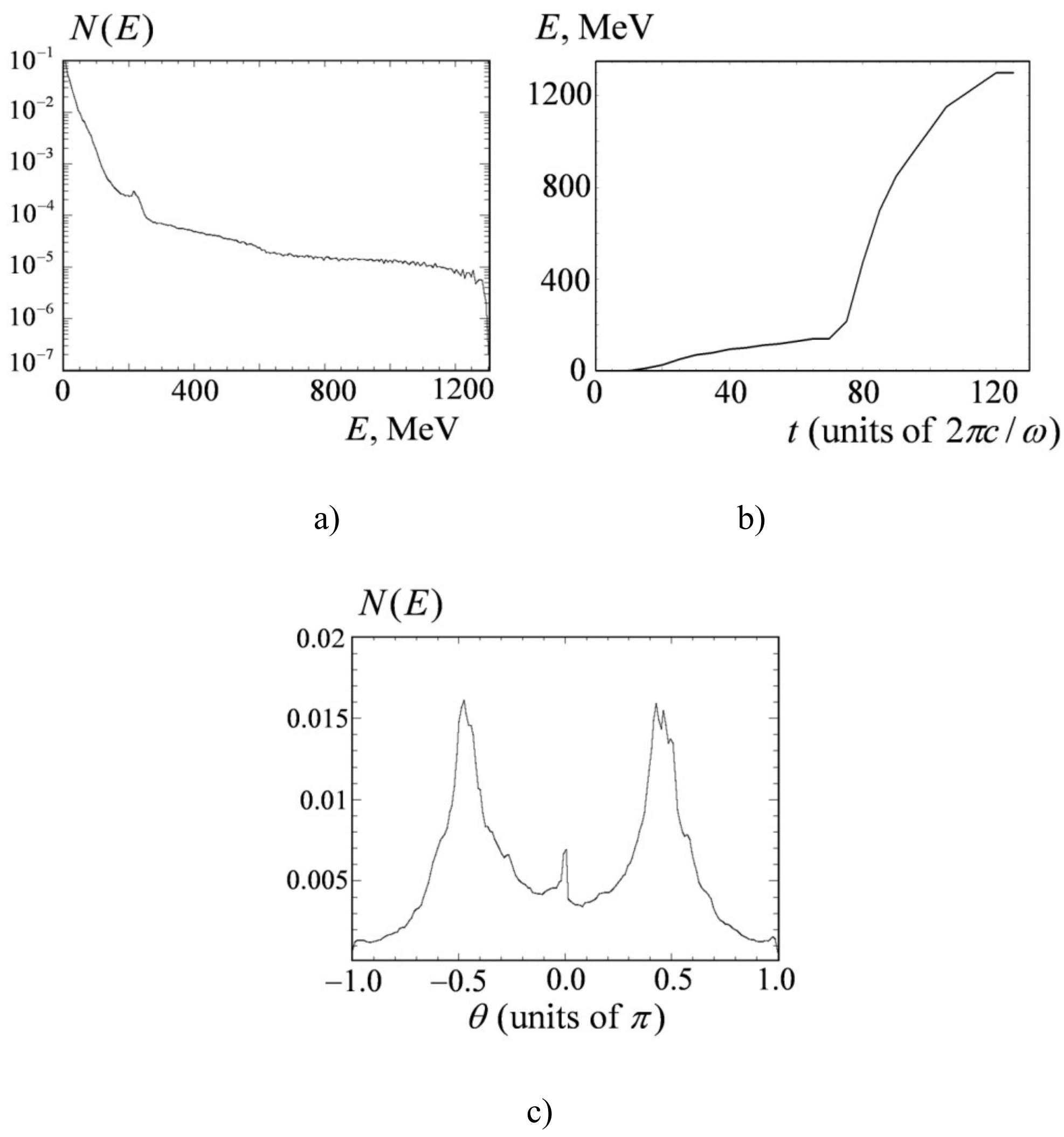

**Fig. 6 a)** The spectrum of protons at t=140; **b)** The dependence of maximum proton energy on time; **c)** Angular distribution of protons, n=3.0 $n_{cr}$, P=1 PW, f/D=1.5, L=50 λ

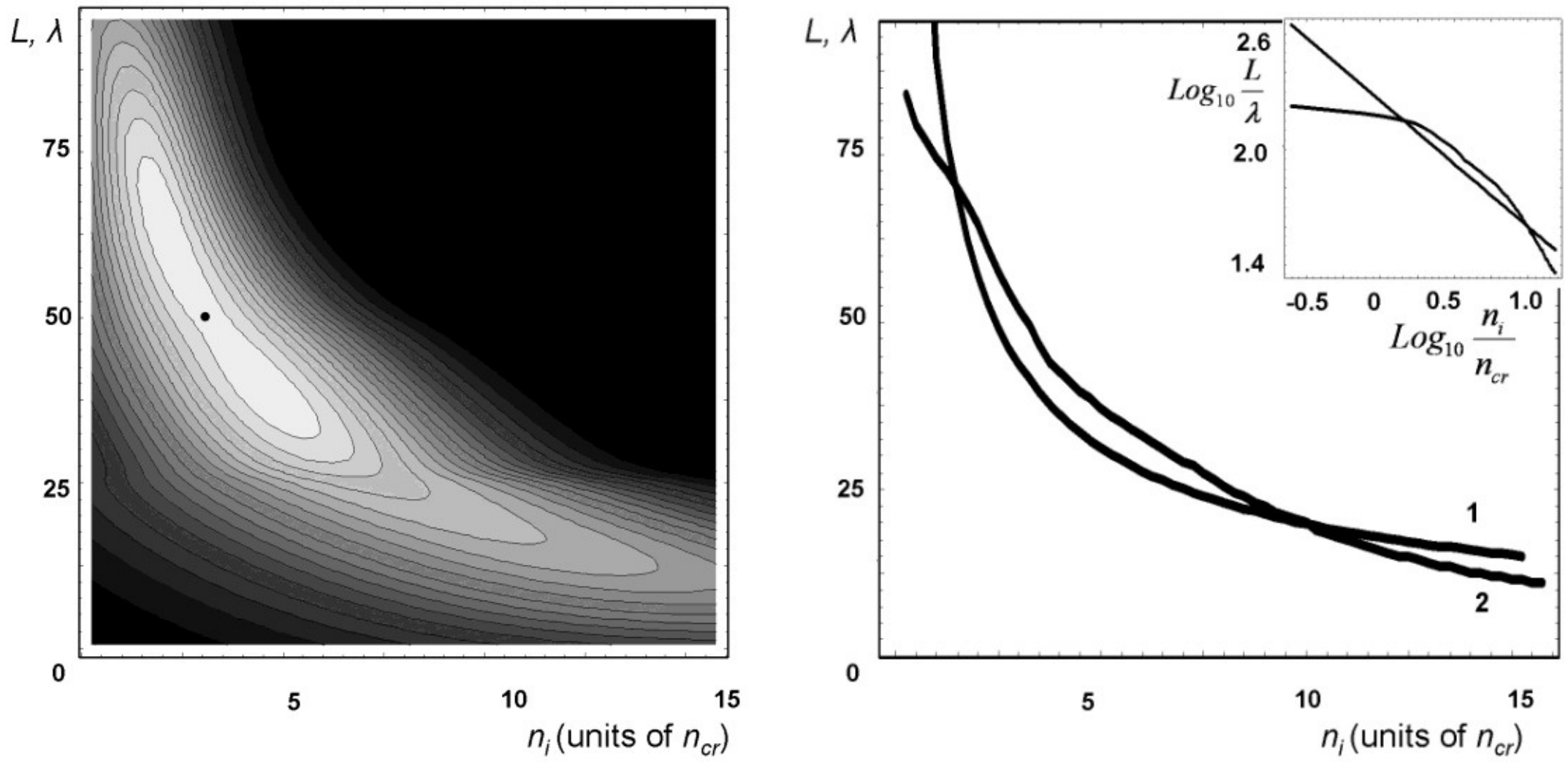

a) b)

**Fig. 7 a)** The distribution of proton maximum energy in plane ($n_e$, $L$); the black dot marks the optimal target for proton acceleration by a 1 PW pulse: $n_e=3.0n_{cr}$ and $L=50\ \lambda$, which gives a maximum energy of 1.3 GeV; **b)** The dependence of target thickness on targetdensity corresponding to maximum accelerated proton energy (2) and its fit by a function $f\ /\ n_e^b$, $f=442,\ b=0.65$, (1); the log-log plot of $L(n_i)$ is shown in the inset.

It is evident that there should be a strong dependence of the acceleration effectiveness on the target thickness and density connected with a variation in energy transfer efficiency from laser pulse to the electrons and protons. This is due to the fact that if the target is very thick the pulse will not penetrate the target and generate an accelerating field, and if the target is too thin then the pulse will transfer a negligible amount of its energy into accelerating fields. For low densities, tight focusing of the pulse leads to filamentation, causing a rapid drop in acceleration efficiency. For high densities the channel radius is reduced, so the generated accelerating fields are also reduced. Hence, the laser pulse properties should be properly matched to the target conditions to ensure optimal conditions for acceleration. By optimal conditions we mean that the laser pulse is able to go through the target without filamentation, and the laser pulse energy is almost all converted into the energy of fast electrons.

That is why we perform a two-parameter scan of the maximum protons energy. In Fig. 7a we show the results of this scan, *i.e.* the dependence of the maximum proton energy on target thickness and density. The parameter space was explored by a large number of simulations to generate the plot. The shape of the surface indicates that for every density there is an optimal target thickness that maximizes the proton energy. Moreover the shape suggests that the optimal target thickness for each value of target density can be determined from the condition $n_e^b L = const$, where $b$ is some number. Fitting the results of 2D PIC simulations we obtain that $b=0.65$ (see Fig. 7b, where the dependence of target thickness on target density for optimal proton acceleration is plotted along with its fit by a function $f / n_e^b$, $f=442,\ b=0.65$) . We should mention here that the data and the power law fit are in good agreement for $1.6<n_i/n_{cr}<10$, as it can be seen from the inset Fig. 7b. The comparison of this scaling with the analytical results and the discussion of the scaling applicability range are carried out in the next section.

According to the results presented in Fig. 7a the maximum proton energy of 1.3 GeV in the interaction of a 1 PW laser pulse with a target of near-critical density is obtained for a target thickness of 50 λ and density of $3.0n_{cr}$.

## III. The optimal thickness and its scaling with interaction parameters.

As it was demonstrated in the previous section there exists an optimal target thickness that maximizes the accelerated proton energy for given laser and target properties. In other words the optimal target thickness corresponds to most efficient transformation of laser pulse energy into the energy of protons, which are accelerated by the longitudinal quasistaic electric field. As it was shown in the previous section this field is generated due to the expansion of the magnetic field as it exits the channel. The magnetic field in its turn is connected with the electron current along the axis of the channel. These are the electrons which are accelerated by the laser pulse in the forward direction. The magnetic field, induced by the electron current, grows as long as these electrons are accelerated. That is why the optimal target thickness should be equal to the electron acceleration length. For the targets of such density, as considered in this paper, the acceleration length is determined by the laser depletion length. In other words the target thickness should be equal to laser depletion length. We can make an estimate of the optimal target thickness from the condition that all the laser energy ($W_p$) is transferred to the energy of electrons ($W_e$), which were initially in the would-be propagation channel: $W_p = W_e$, where $W_e = \pi R^2 L_{ch} n_e a m_e c^2$, $R$ is the radius of the channel, $L_{ch}$ is the channel length, and $a$ is the maximum value of the laser pulse dimensionless vector-potential in the channel. Here we assumed that the electrons acquire an average energy of $a m_e c^2$ after being pushed out from the channel in the transverse direction. We also assume that the thickness of the target is much larger than the pulse length, so the channel can be established. The energy of the pulse can be estimated as follows: $W_p = \pi R^2 \tau a^2 m_e c n_{cr}$ .Then

$$a \sim \frac{n_e}{n_{cr}} \frac{L_{ch}}{L_p}, \qquad (1)$$

where $L_p = \tau / c$ is the length of the laser pulse. If we assume that the diameter of the plasma channel is about $R = a^{1/2} c / \omega_{pe} / 2 = \lambda (n_e / n_{cr})^{1/2} a^{1/2} / 2$, and express $a$ in terms of laser pulse energy then $n_e^{2/3} L_{ch} = const$. This relation agrees well with the results of 2D PIC simulations, which gives a power of 0.65 for density in the scaling (see Fig. 7b).

Let us utilize a simple model to more carefully estimate a constant of proportionality in Eq. (1). In order to do this we need to calculate with accuracy better then above the energy of the laser pulse inside the self-generated channel. We will also use the results of 2D PIC simulations below to prove the assumption that the average energy of electrons that are ejected from the channel is $am_e c^2$. Let us first calculate the energy of the laser pulse. Since we are considering the interaction of an intense laser pulse with a plasma of near-critical density it is plausible to expect that the walls of the self-generated channel will have high density. The laser pulse will be contained inside the channel almost completely since it would not be able to penetrate these high density walls. That is why in order to estimate the energy of the laser pulse inside the channel we can use a well-known result for the behavior of the EM wave inside a waveguide [25]. In doing so, we neglect the laser pulse energy loss, while the channel is being established. Since initially the pulse has no longitudinal component of electric field and it is tightly focused on the front surface of the target it is reasonable to assume that the mode with lowest transverse frequency will propagate through this self-generated plasma waveguide. This will be an H-wave ($E_x=0$) with

$$H_x = A\, J_1(\kappa r) \cos(\omega t - kx)\,, \tag{2}$$

where $J_1$ is the Bessel function of the first kind and $\kappa = 1.84/R$, here $R$ is the radius of the waveguide. The transverse components of electric and magnetic field are expressed through the longitudinal one according to well-known formulae

$$E_r = \frac{i\omega}{\kappa^2 c}\frac{\partial H_x}{\partial r}, \quad H_r = \frac{\omega}{\kappa^2 c}\frac{\partial H_x}{\partial r}\,. \tag{3}$$

Since the amplitude of the transverse field inside the waveguide is $E_0 = mc\omega a / e$ then the amplitude of the magnetic field in Eq. (2) is $A = (2c\kappa / i\omega)(mc\omega a / e)$. The energy of the EM wave traveling in such a waveguide per unit length is

$$w = \frac{\omega^2}{8\pi\kappa^2 c^2} \int |H_x|^2 df = \frac{\pi R^2}{2} n_{cr} a^2 m_e c^2 \left[ \left( J_1(\kappa R) \right)^2 - J_0(\kappa R) J_2(\kappa R) \right] \tag{4}$$

If we take into account the finite duration of the pulse and assume that it is Gaussian, then the integration over time will produce the following result

$$\int_{-\infty}^{+\infty} \left[ \exp\left( -\frac{4t^2}{\tau^2} \right) \right]^2 dt = \sqrt{\frac{\pi}{8}} \tau , \tag{5}$$

Then the energy of the laser pulse inside the self-generated plasma waveguide is given by the following formula

$$W_p = \pi R^2 \tau a^2 m_e c n_{cr} K , \tag{6}$$

where $K = \sqrt{\pi / 32} \left( J_1(\kappa R)^2 - J_0(\kappa R) J_2(\kappa R) \right)$. Above we assumed that the average energy of electrons accelerated by the laser pulse is $a m_e c^2$, and the total energy of these electrons is

$$W_e = \pi R^2 L_{ch} n_e a m_e c^2 , \tag{7}$$

Let us test this assumption against the results of 2D PIC simulations. The typical spectrum of electrons is presented in Fig. 8. It is obvious that the electron energy does not follow the scaling $a^2$(*=40000,* or *20* GeV electron energy)*,* since the maximum electron energy is equal to 1 GeV. Let us estimate the average energy of accelerated electrons. We assume that all the energetic electrons comes from the volume occupied by the laser produced channel, and define their number as $N_{ch}$. Then from the electron spectrum we can determine the threshold energy, $E_{th}$, *i.e.* the minimum energy that the electron initially from the channel can have:

$$N_{th} = \int_{E_{th}}^{E_{\max}} N(E) dE \tag{8}$$

The average energy of these electrons then can be defined as

$$\bar{E} = \frac{1}{N_{ch}} \int_{E_{th}}^{E_{max}} EN(E)dE \tag{9}$$

For the spectrum, shown in Fig. 8, the average energy is $\bar{E} = 93\,\text{MeV}$. This means that if $\bar{E} = am_ec^2$, as we assumed, then $a = 186$. Such value of $a$ coincides well with the value obtained in 2D PIC simulations for the amplitude of vector-potential in the propagation channel. This supports the assumption that the bulk electrons which are accelerated by the laser gain energy that scales as $a$.

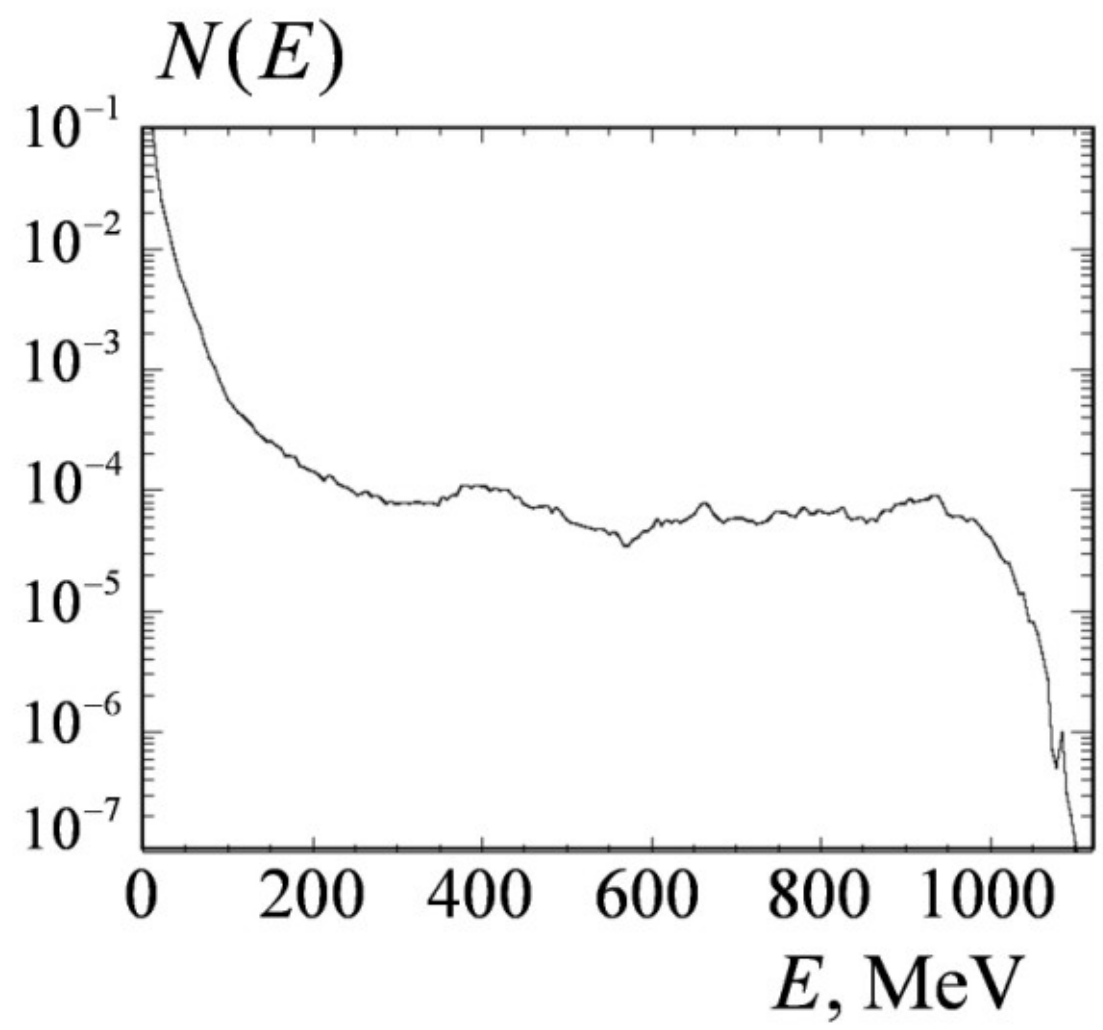

**Fig. 8.** The spectrum of electrons for a 1 PW laser pulse interacting with a 50 λ thick target with density of 3.0 $n_{cr}$.

Then

$$L_{ch} = a\frac{n_{cr}}{n_e}L_pK \tag{10}$$

or

$$a = \frac{1}{K}\frac{n_e}{n_{cr}}\frac{L_{ch}}{L_p}.$$

If we consider a propagation of the electromagnetic pulse in a 2D waveguide then in the above calculation only factor K will change, which is directly connected to the dimension of the problem and exact form of the laser field in the waveguide. In 2D case $K=1/10$ (instead of $K=1/13.5$ in 3D). For the parameters, used in 2D PIC simulations: the maximum value of the vector potential inside the channel $a=200$, $n_e=3.0n_{cr}$, and $L_p=10\lambda$, the optimal target thickness is $L_{ch}=50\lambda$, which is in good agreement with the prediction of condition (10).

The scaling (10) can be rewritten in terms of laser pulse energy. Taking into account that the radius of the channel is $R=\lambda/2\left(n_e/n_{cr}\right)^{1/2}a^{1/2}$ and that $a$ can be determined from the expression for the energy of the laser pulse, we obtain

$$\frac{n_e^{2/3}L_{ch}}{W_p}=C, \tag{11}$$

where the constant $C$ is

$$C=\left(\frac{4K^2 n_{cr}L_p^2}{\pi\lambda^2 m_e c^2}\right)^{1/3}$$

As we can see from the comparison of results of 2D PIC simulation and the simple analytical model the scaling fails at low and high densities, while it works well for the intermediate ones ($1.6<n_i/n_{cr}<10$, see Fig. 7b). In the case of low densities, $n_i<1.6n_{cr}$, the filamentation of the laser pulse and Langmuir wave generation due to self-modulation decrease the optimal target thickness by adding new sources of laser pulse energy depletion not accounted for in the scaling (11). In the case of high densities, $n_i>10n_{cr}$, other mechanisms of ion acceleration come into play and their effectiveness is no longer determined by the depletion of the laser pulse. Moreover the optimal target thickness that follow from the result of 2D PIC simulations for high density targets is no longer much larger than the pulse length, whereas the condition of the theoretical model applicability is $L_{ch}$ □ $_{r}$ .

In contrast to the results of the present paper the scaling of the optimal target thickness for ion acceleration from thin foils is determined from the condition of the relativistic transparency of the foil. This condition stems out of the requirement that the electric field of the laser is equal to the electric field of the ion core, stripped of all the electrons, and which is as follows [26]:

$$a = \pi \frac{n_e}{n_{cr}} \frac{l}{\lambda}, \tag{12}$$

here $l$ is the foil thickness and λ is the laser wavelength. This scaling was thoroughly studied in Ref. [27] in a series of 2D PIC simulations, where the ion acceleration from thin double-layer foils was considered. The difference in scaling is due to different targets and thus to different interaction properties. The effect of relativistic transparency on enhanced ion acceleration schemes was studied experimentally in Ref. [28]

The maximum energy of the accelerated protons can be estimated from the fact that the acceleration itself takes place in the region which is of order of channel diameter. The electric field, which accelerated protons, should be of the order of the magnetic field generated by the accelerated electron bunch, i.e. $E \approx B \approx 4\pi e n_e \gamma_e^2 R_{e,b}$ , where $R_{e,b} << a^{1/2} c / \omega_{pe}$ is the transverse size of the electron bunch. Here we take into attention known fact that the relativistic electron bunch density at the equilibrium due to the plasma lensing effect is in a factor $\gamma_e^2$ higher than the background plasma density $n_e$. Then the proton energy is given by

$$E_p \approx e^2 n_e \gamma_e^2 (a^{1/2} c/\omega_{pe})^2 \approx a\gamma_e^2 m_e c^2 \tag{13}$$

The coefficient of proportionality should be determined from PIC simulations.

It is easy to find a relationship between the dimensionless amplitude $a$ of laser beam inside the self focusing channel of the radius $a^{1/2} c/\omega_{pe}$ and the laser power $P$. It reads

$$a = \left[8\pi (P/P_c)(n_e/n_{cr})\right]^{1/3}, \tag{14}$$

where $P_c = 2m_e^2 c^5/e^2$ =17 GW. For P=1 PW and $n_e/n_{cr}$ =3 it yields $a \approx 10^2$. As we can see, Eq. (13) gives 1 GeV proton energy for 1 PW laser if the bulk energy of fast electrons corresponds to $\gamma_e \approx 7$.

## IV. Conclusions

In this paper we have studied, using 2D PIC simulations, the mechanism of proton acceleration from near-critical density plasma. In this scheme the laser pulse burns through the target generating strong electric and magnetic fields in the propagation channel. The magnetic field begins to expand along the transverse direction upon exiting the channel, pushing the electron component of plasma inside the target and generating the quasistatic electric field. This electric field accelerates and collimates ions from the thin filament which is formed in the propagation channel.

The results of simulations indicate the existence of an optimal target thickness that maximizes the energy of accelerated ions. Since the quasistatic fields are generated by the electron current the optimal target thickness is connected with the maximum energy transfer from the laser pulse to the plasma electrons. A parameter scan using 2D PIC simulations confirmed a scaling law for proton acceleration from the near critical density targets, given by:

$$L_{ch} = a \frac{n_{cr}}{n_e} L_p K, \tag{15a}$$

or

$$a = \frac{1}{K} \frac{n_e}{n_{cr}} \frac{L_{ch}}{L_p} \tag{15b}$$

or in terms of laser pulse energy

$$\frac{n_e^{2/3} L_{ch}}{W_p} = \text{constant} \tag{15c}$$

where $K$ is a numerical factor determined by the laser pulse profile and its propagation inside the self-generated plasma channel. This scaling comes from the requirement of optimal laser pulse depletion in plasma and the fact that the laser pulse should be able to burn through the target. Comparing the analytical scaling and the results of 2D PIC simulations we came to a conclusion that the scaling (11) has an applicability range of $1.6 < n_i/n_{cr} < 10$. For high densities ($n_i > 10n_{cr}$) other mechanisms of ion acceleration come into play and their effectiveness does not depend on the laser pulse energy depletion. Thus the scaling (11) is no longer valid. For low densities ($n_i < 1.6n_{cr}$) the scaling fails due to the fact that the filamentation of the laser pulse and Langmuir wave generation due to self-modulation decrease the optimal target thickness by adding new sources of laser pulse energy depletion not accounted for.

A parameter scan over different values of target density and thickness allowed us to determine the optimal conditions for proton acceleration by a 1 PW laser pulse focused to a 1 μm focal spot from a near-critical density target: a thickness of 50λ and density of 3.0 $n_{cr}$.

The resulting protons have the maximum energy of 1.3 GeV. This mechanism also works for lower intensities. For example, a 100 TW laser pulse with intensity of $6x10^{21}$ W/cm$^2$ interacting with 2.7$n_{cr}$ dense plasma produces 40 MeV maximum energy protons. For a 500 TW laser pulse the maximum proton energy goes up to 770 MeV.

Let us briefly mention that the protons accelerated by the studied in this paper mechanism can be of interest for applications in hadron therapy. For a 225 TW laser pulse which interacts with a 2.25 $n_{cr}$, 60 $\lambda$ plasma slab the maximum proton energy is about 300 MeV. The number of protons accelerated to the energy of 250 MeV with an energy spread of 1% is estimated to be about $10^8$. As a result these proton beam parameters make this acceleration regime an interesting potential candidate for proton therapy.

## Acknowledgements

This work was supported by the National Science Foundation through the Frontiers in Optical and Coherent Ultrafast Science Center at the University of Michigan and by grant (R21 CA120262-01) from the National Institute of Health and the International Science and Technology Center (project 2289). The authors would like to acknowledge fruitful discussions with A. Brantov.